\documentstyle[aps,prd,12pt]{revtex}
\def\lsim{\lesssim}

\begin{document}
\pagestyle{empty}
\begin{center}
{\Large \bf Planetary Perturbation with cosmological Constant}
\end{center}
\vspace {0mm}
\begin{center}
 Ishwaree P. Neupane \footnote{Electronic mail: ishwaree@phya.snu.ac.kr}\\
{Department of Physics,Seoul National University, Seoul 151-742, Korea}
\footnote{ On leave from {\it Department of Physics, Tribhuvan University, Kirtipur, Kathmandu, Nepal}} 
\end{center}
\begin{abstract}
A contribution of quantum vacuum to the energy momentum tensor is inevitably experienced in the present universe. One requires the presence of non-zero cosmological constant ($\Lambda$) to make the various observations consistent. A case of $\Lambda$ in the Schwarzschild de Sitter space-time shows that precession of perihelion orbit provides a sensative solar test for non-zero $\Lambda$. Application of the relations involving $\Lambda$ to the planetery perturbation indicates the values near to the present bound on $\Lambda$. Also suggested are some relations in vacuum dominated flat universe with a positive $\Lambda$. 
\end{abstract}
The cosmological constant[1-3] has been an outstanding problem in various microscopic theories of particle physics and gravity for the past several decades, ever since Einstein introduced it in the field equations to avoid an expanding 
universe. The standard model of cosmology based on the ideas arising from 
particle physics involves the following trilogy of ideas: (i) $\Omega$=1, (ii) $\Lambda=0$ and (iii) $\Omega_{matter}\lsim 0.9$[4]. But, in reference to the  large scale structure measurements, the density of the matter insufficient to  result in a flat universe($\Omega=1$) suggests a non-zero $\Lambda$. Now one 
would prefer either (1) $\Omega\neq1$, $\Lambda=0$, $\Omega_{matter}\approx 0.2-0.4$ or (2) $\Omega\equiv1$, $\Lambda\neq0$, $\Omega_{matter}\approx 0.2-0.4$.
A small non-vanishing $\Lambda$ is also required to make the two independent 
observations: the Hubble constant,$H_o$ which explains the expansion rate of 
the present universe, and the present age of the universe $(t_o)$ consistent 
each other[5]. This has forced us critically re-examine the simplest and most 
appealing cosmological model- a flat universe with $\Lambda=0$[4]. A flat 
universe with $\Omega_m\equiv0.3$ and $\Omega_m+\Omega_{\Lambda}=1$ is most
preferable and $\Lambda=0$ flat universe is almost ruled out[4,5]. Indeed,
$\Lambda$ follows from the dynamical evolution of our universe when one 
interprets it as the vacuum energy density of the quantized fields. The large
scale structure measurements of the present universe would imply $\Lambda$ to 
have an incredibly small value $\lsim10^{-47} (GeV)^4$[2-5], while the quantum
field theories in curved spacetime predict quite different values of vacuum
energy density\footnote{in units $8\pi G=c=1$, we denote $\rho_v$ by $\Lambda$},($\rho_v$) in the early universe. In particular, $\Lambda_{GUT}\sim10^{64}(Gev)^4$, $\Lambda_{EW} \sim 10^8 (GeV)^4$ and $\Lambda_{QCD} \sim10^{-4} (GeV)^4$[3]. This is the source of cosmological constant problem.\\

In this letter we consider a case of non zero $\Lambda$ in Schwarzschild de Sitter space-time and study its effect on the geodetic motions of the planets in vacuum dominated universe.\\

To the present limit $\Lambda_{o} \lsim 10^{-47}(GeV)^4$ arising from the large scale structure measurements, there might exist its correspondence to the  limit $\Lambda_o \lsim 10^{-120} M_{pl}^{2}$ in natural units[6], which is $\sim10^{-55} cm^{-2}$; a value consistent with that of particle data group \footnote{\it{Review of Particle Physics}, Eur.Phys.J.C3 (1998) 70}i.e., $0\leq|\Lambda_o|\leq 2.2\times 10^{-56}cm^{-2}$.\\

The vacuum expectation value of energy momentum tensor of quantum fields in  de Sitter space[7,2] takes the form $<T_{\mu\nu}^{vac}>=\rho_{vac} g_{\mu\nu}$. So, a model universe with an additional term $\rho g_{\mu\nu}$ in the Einstein field equation is highly motivated and $\Lambda$ corresponding to the vacuum energy density enters in the form
\begin{eqnarray}
R_{\mu\nu} - \frac{1}{2}g_{\mu\nu}R =G_{\mu\nu}= 8\pi GT_{\mu\nu} - \Lambda g_{\mu\nu}
\end{eqnarray}
A generally spherically symmetric metric is described by the form
\begin{eqnarray}
d\tau^2 = e^{2\alpha(r,t)} dt^2 - e^{2\beta(r,t)} dr^2 - r^2 (d\theta^2 + \sin^2 d\varphi^2)
\end{eqnarray}
where $\alpha$ and $\beta$ are some functions of $(r,t)$. Corresponding to the field equations $G_{\mu\nu}=-\Lambda g_{\mu\nu}$, the generalized and spherically symmetric vacuum solution[8] for the above metric by allowing non-zero cosmological constant (in units $c=1$) is given by
\begin{eqnarray}
d\tau^2=B(r)dt^2-A(r)dr^2-r^2(d\theta^2+\sin^2\theta d\varphi^2)
\end{eqnarray} 
where $B(r) = A(r)^{-1} = 1-2GM/r-\Lambda r^2/3$. This metric is considered as the Schwarzschild de-Sitter metric and hence the space determined by it is not asymtotically flat as the case in Schwarzschild metric, for $\Lambda$ related 
to the vacuum energy density implies a pre-existing curvature[8]. It is easy to see that the Lagrangian and Hamiltonian for this metric are equal and hence no potential energy is involved in the problem. By rescaling $\tau$ and setting 
$\theta=\pi/2$ ( i.e., an equatorial plane), we get
\begin{eqnarray}
E^2B(r)^{-1}-A(r)\dot{r}^2-\frac{J^2}{r^2}={\cal L}= +1 ~\mbox{or} ~0
\end{eqnarray}
for the time like or null geodesics respectively; where $E=(1-2MG/r-\Lambda 
r^2/3)\dot{t}$ and $J=r^2 \dot{\varphi}$ are the constants related to the 
energy and momentum of the test particle respectively. Here dot represents differentiation w.r.t. the affine parameter, $\tau$. For time like geodesic, considering $r$ as a function of $\varphi$, we get
\begin{eqnarray}
\frac{A(r)}{r^4} \Big(\frac{dr}{d\varphi}\Big)^2 + \Big(\frac{1}{J^2}+\frac{1}{r^2}\Big) = \frac{E^2}{B(r) J^2}
\end{eqnarray}
The solution of this orbit equation is determined by a quadrature
\begin{eqnarray}
\varphi=\pm\int A(r)^{1/2} r^{-2} \Big[ \frac{E^2}{B(r)J^2}-\frac{1}{J^2}-\frac{1}{r^2} \Big]^{-1/2}dr
\end{eqnarray}
Defining $r_-$ and $r_+$ as perihelion and aphelion of a closed elliptic orbit, the angular orbit precession in each revolution is $\Delta\varphi = 2|\varphi(r_+) - \varphi(r_-)| - 2\pi$.
Following the treatment given by Weinberg[9] with a slightly different technique to evaluate some constants, a solution valid for slightly eccentric orbit gives a precession
\begin{eqnarray}
\Delta\varphi=\frac{3\pi r_s}{L}+\frac{2\pi\Lambda L^3}{r_s}+\frac{2\pi\Lambda L r_s}{3} +\dots
\end{eqnarray}
where $L$ is the semilatus rectum and $r_s = 2GM/c^2$ the Schwarzschild radius.The first term is the same as general relativity prediction for the precession of perihelion orbit obtained without introducing cosmological constant in the metric(3) and gives the precession for inner planets very much consistent with the experimental observation. Evidently, the extra precession factor obtained by introducing a positive cosmological constant is therefore
\begin{eqnarray}
\Delta\varphi_{\Lambda} = \frac{2\pi\Lambda L}{r_s} \Big( L^2+\frac{r_{s}^{2}}{3}+\dots \Big)
\end{eqnarray}
For planetary system, the contribution from second and higher terms is negligible compared to the first term.One can see that the contribution of the second term in eqn(8) would be significant only for very high eccentric orbits and large Schwarzschild radius. So for very massive binary star systems such as Great Attractor(GA) and Virgo Cluster with highly eccentric orbits, the value of cosmological constant may show up. So the main effect of the term involving $\Lambda$ in eqn(7) is to cause an extra additional advance of the perihelion of the orbit by an amount 
\begin{eqnarray}
\Delta\varphi_{\Lambda}\equiv\frac{2\pi\Lambda L^3}{r_s}= \frac{\pi\Lambda c^2 a^3 (1-e^2)^3}{GM}
\end{eqnarray}
where $a$ is semimajor axis and $e$ eccentricity of the orbit. In planetary
motion the accuracy of precession of the orbit degrades rapidly as we move away from the sun mainly by two reasons: for smaller eccentricity the observation of the perihelia becomes more uncertain and also as $L$ increases the precession per revolution decreases.\\

In the case of Mercury, the extra precession factor $\Delta\varphi_{\Lambda}\approx 0.1''$ per century (i.e.,the maximum
 uncertainty in the precession of the perihelion) would imply
 $\Lambda\leq 3.2\times 10^{-43} cm^{-2}$. With the value of 
$|\Lambda|\leq 10^{-56} cm^{-2}$, for Mercury, one gets  
$\Delta\phi_{\Lambda}\leq 3.6\times 10^{-23}$ arc second per revolution; which is unmeasurably small and very far from the present detectable limit of VLBI
 i.e.$3\times 10^{-4}$arc second. With $\Lambda\leq 10^{-56} cm^{-2}$, for
 Pluto with $L= 5.5\times 10^{14}$ cm, one gets  $\Delta\varphi_{\Lambda}=3.5\times 10^{-17}$ arc second per revolution; which is also unmeasurably small.
 For Pluto with $\Delta\phi_{\Lambda}\leq 0.1''$ per revolution, one gets
 $\Lambda\leq 3.3\times10^{-49}cm^{-2}$, which is near to the present bound on cosmological constant i.e., $0\leq|\Lambda_o|\leq 2.2\times10^{-56} cm^{-2}$.\\

For the case of bound orbits, a relation between the cosmological constant and the minimum orbit radius can be expressed by $r_{min}=(3MG/\Lambda c^2)^{1/3}$. This suggests that the effect of $\Lambda$ can be expected to be significant only at large radii. Also from eqn(9) it seems more reasonable to argue that more distant planets would give better limit to cosmological constant. For circular orbits one can generalise the relation further, i.e. $\Delta\phi_\Lambda=\pi\Lambda c^2 a^3/GM$. If we define $\rho$ as the average density within a sphere of radius $a$ and $\rho_{vac}=\Lambda c^2/8\pi G$ as the vacuum density equivalent of the cosmological constant, one gets $\Delta\varphi_{\Lambda} = 6\pi(\rho_{vac}/\rho) $ radians/revolution. If we evaluate the value at $r\equiv L$, we get
\begin{eqnarray}
\Delta\varphi_{\Lambda}=\frac{\pi\Lambda c^2 r^3}{GM}=\frac{3P^2H_{o}^{2}\Omega_{vac}}{4\pi}
\end{eqnarray}
where $P=(2\pi r^3/ G M)^{1/2}$ is the period of revolution, $H_o$ the present value of Hubble constant and $\Omega_{vac}=\rho_{vac}/\rho_c$ the vacuum density parameter with $\rho_c = 3H_o^2/8\pi G$ and $\rho_{vac} = \Lambda c^2/ 8\pi G$.\\  

The microscopic theories of particle physics and gravity suggest a large contribution of vacuum energy to energy momentum tensor. However, all cosmological observations to date show that $\Lambda$ is very small and positive. It is logical to argue that an extremely small value $\Lambda$ makes us unable to measure the extra precession with the required precision. It is here worthnoting that $\Lambda$ must be quite larger than $10^{-50}cm^{-2}$ to observe its effects possibly with an advance of additional precession of perihelion orbit in the inner planets. It judges more sound to argue that only the tests based on large scale structure measurements of the universe can put a strong limit on $\Lambda$. Nevertheless, the precession in the perihelia of the planets provides a sensitive solar test for a cosmological constant. Whether a non-zero Cosmological constant exists is one of the hot issues in various theories of particle physics and gravity. But it is certain that the planetary perturbations cannot be used to limit the present value of cosmological constant.\\

{\large\bf References}
\begin{description}
\item{[1]} W.A.Hiscock, {\it Phys.Lett.}166B, \underline{Vol.3} (1986) 285
\item{[2]} M.Ozer and M.O.Taha, {\it Phys.Lett.} 171B (1986) 363
\item{[3]} S.Weinberg, {\it Rev.Mod.Phy.} 61(1989)1
\item{[4]} L.M.Krauss and M.S.Turner, {\it J.Gen.Rel.Grav.}, 27 (1995) 1137
\item{[5]} Anup Singh, {\it Phys.Rev.} D52 (1995) 6700
\item{[6]} J.Lopez and D.Nanopoulos, {\it Mod.Phys.Lett.} A11 (1996) 1
\item{[7]} L.Ford, {\it Phys.Rev.} D28(1983) 710
\item{[8]} G.Gibbons and S.Hawking, {\it Phys.Rev.} D15 (1977) 2738
\item{[9]} S.Weinberg, {\it Gravitation and Cosmology}, Wiley Publ.NewYork (1972)
\end{description}
\end{document}